\documentclass[runningheads]{llncs}
\usepackage{graphicx}
\usepackage{epsfig}
\usepackage{amsmath}
\usepackage{amssymb}
\usepackage{bbding}
\usepackage{amsfonts}
\usepackage{multirow}
\usepackage{booktabs}    
\usepackage{xcolor}
\usepackage{hyperref}

\newcommand{\real}{\mathbb{R}}

\newcommand{\figref}[1]{Fig.~\ref{#1}}
\newcommand{\tabref}[1]{Table~\ref{#1}}
\newcommand{\secref}[1]{Sec.~\ref{#1}}

\begin{document}
\title{
Small Lesion Segmentation in Brain MRIs with Subpixel Embedding
}
%
%

\renewcommand\footnotemark{}
\renewcommand\footnoterule{}

\author{
Alex Wong$^{\dagger, 1 (}$\Envelope$^)$, Allison Chen$^{\dagger, 1}$, Yangchao Wu$^1$, Safa Cicek$^1$, \\
Alexandre Tiard$^1$, Byung-Woo Hong$^2$, and Stefano Soatto$^1$
\thanks{$^\dagger$ denotes authors with equal contributions.}
}
\authorrunning{A. Wong et al.}
%
\institute{
$^1$University of California, Los Angeles, CA, USA \\
\email{alexw@cs.ucla.edu} \\
$^2$Chung-Ang University, Seoul, Korea
}
\maketitle              
\begin{abstract}
We present a method to segment MRI scans of the human brain into ischemic stroke lesion and normal tissues. We propose a neural network architecture in the form of a standard encoder-decoder where predictions are guided by a spatial expansion embedding network. Our embedding network learns features that can resolve detailed structures in the brain without the need for high-resolution training images, which are often unavailable and expensive to acquire. Alternatively, the encoder-decoder learns global structures by means of striding and max pooling. Our embedding network complements the encoder-decoder architecture by guiding the decoder with fine-grained details lost to spatial downsampling during the encoder stage. Unlike previous works, our decoder outputs at $2\times$ the input resolution, where a single pixel in the input resolution is predicted by four neighboring subpixels in our output. To obtain the output at the original scale, we propose a learnable downsampler (as opposed to hand-crafted ones e.g. bilinear) that combines subpixel predictions. Our approach improves the baseline architecture by $\approx 11.7\%$ and achieves the state of the art on the ATLAS public benchmark dataset with a smaller memory footprint and faster runtime than the best competing method. Our source code has been made available at: \\ \url{ https://github.com/alexklwong/subpixel-embedding-segmentation}.
\end{abstract}

\section{Introduction}

\def\netname{SPiN }

A stroke occurs when a lack of blood flow prevents brain tissue from receiving adequate oxygen and nutrients. This condition affects over 795,000 people annually \cite{virani2020heart}. The severity of the outcome, including disability and paralysis, depends on the location and intensity of the stroke, as well as the time of diagnosis \cite{atlantis2004association,van2007acute}. Preserving cognitive and motor functions, therefore, hinges on localizing stroke lesions quickly and precisely. However, doing so manually requires expert knowledge, is time consuming, and is ultimately subjective \cite{liew2018large,maier2017isles}.

We focus on automatically segmenting ischemic stroke lesions, which account for 87\% of all strokes \cite{virani2020heart}, from T1-weighted  anatomical magnetic resonance imaging (MRI) brain scans. These lesions are characterized by high variability in location, shape, and size -- the latter two are problematic for conventional convolutional neural networks (CNNs) where precision of irregularly shaped lesion boundaries and recall of small lesions are critical measures of success. Due to aggressive spatial downsampling (i.e. max pooling, strided convolutions) customary in CNNs, details of local structures are lost in the process. Yet, the spatial downsampling is necessary for obtaining a global representation of the input while using fixed-size filters with limited receptive fields. 
The outcome of which are segmentations with ambiguous boundaries between lesion and normal tissues and missed lesions that occupy small number of voxels in the MRIs.

We propose to retain small local structures by learning an embedding that maps the input to high dimensional feature maps of \textit{twice} the input resolution. Unlike the typical CNN, we do not perform lossy downsampling on this representation; hence, the embedding preserves local structures, but lacks global context. When combined with the standard encoder-decoder e.g. U-Net \cite{ronneberger2015u}, the embedding complements the encoder-decoder by supplying the decoder with fine-grained detail information to guide segmentation. Our network also outputs at twice the resolution of the input, representing each element in the input with a $2 \times 2$ neighborhood of predictions. The final output is obtained by combining the four predictions (akin to an ensemble) as a weighted sum where the contribution of each prediction is learned from the data. Our design not only enables the network to produce robust segmentations but also localize small lesions (\figref{fig:results_atlas_small_lesions}).

\textbf{Our contributions} include (i) an embedding function that preserves fine-grained details of the input by mapping it to larger spatial dimensions, (ii) a neural network architecture that leverages the complementary strengths of the proposed embedding and an encoder-decoder to produce predictions at twice the input resolution, and (iii) a learnable downsampler that combines local predictions in an ensemble fashion to yield robust segmentations at the input resolution. Our approach improves the baseline U-Net architecture by $\approx11.7\%$ and achieves the state of the art on the ATLAS \cite{liew2018large,liew2017anatomical} dataset with lower computational burden than the best competing method.

\section{Related Work}
\label{sec:related_work}
\textbf{Lesion Segmentation.} Early works \cite{ciresan2012deep} aggregated classification results for the center pixel of patches sampled from an image. However, \cite{ciresan2012deep} lacked global context, so \cite{Seyedhosseini_2013_ICCV} addressed this with multi-stage cascaded hierarchical models. More recent works build upon the U-Net \cite{ronneberger2015u}, a 2D fully-convolutional network with skip connections and up-convolutions. For example,  \cite{manvel2019radiologist} used a Dual Path Network \cite{chen2017dual} encoder while \cite{tureckova2018isles} leveraged dilated convolutions to inexpensively increase receptive fields. Furthermore, \cite{asadi2020multi} fused the U-net with other high-performing modules, the BConvLSTM \cite{song2018pyramid} and the SENet \cite{hu2018squeeze}, and \cite{qi2019x} introduced X-blocks to the U-Net, leveraging depthwise separable convolutions to reduce computational load. \cite{yang2019clci} used skip connections between successive encoder resolutions to prevent the loss of features and ConvLSTM \cite{shi2015convolutional} modules to maintain localization. 

Recent works also leveraged 3D architectural backbones to improve localization. \cite{zhou2019d} performed 3D convolutions on a subsection of the scan and fused the results with 2D convolutions. \cite{hui2020partitioning} proposed an attention gate to combine 2D segmentations along the axial, sagittal, coronal planes into a 3D volume. However, these works use significantly larger memory footprints and 3D convolutions are computationally expensive -- limiting the models' practicality. We note that while conventional architectures perform well globally (i.e. recovering the coarse shape of lesions) they struggle to segment small lesions that blend into the background. 

\textbf{Super-Resolution.} There is an abundance of works in natural images super-resolution \cite{dong2014learning,dong2015image,shi2016real,tang2020srda,wang2020dual} and a growing number in medical imaging. \cite{rueda2013single} proposed to map MRI images from low to high-resolution with an overcomplete dictionary. \cite{pham2017brain} leveraged SRCNN \cite{dong2014learning} for super-resolving 2D MRI images and fused them to obtain a 3D volume. \cite{pham2019multiscale} handled arbitrary scaling factors with a 3D architecture for multi-modal 3D data. However, these works require low and high-resolution image pairs for training and are limited to the super-resolution task while our method does not rely on a larger resolution ground truth. More recently, \cite{valanarasu2020kiu} introduced Kite-Net, an upsampling encoder that outputs a latent at $8 \times$ resolution followed by a max-pooling decoder to downsample back to the original resolution. Kite-Net is used in parallel with a U-Net for lesion segmentation. Our approach draws inspiration from super resolution and latent over-representations as methods to retain local structure that are often lost in spatial downsampling. However, unlike \cite{valanarasu2020kiu}, we avoid downsampling the latent with pooling (which discards information), and instead employ lossless space-to-depth and depth-to-space \cite{shi2016real} operations to retain fine-grained details. Furthermore, we propose to learn a subpixel embedding at $2 \times$ the original resolution to guide our segmentation, which uses a much smaller memory footprint than \cite{valanarasu2020kiu}. We show that our approach can capture small lesions that are missed by \cite{qi2019x,ronneberger2015u,valanarasu2020kiu,yang2019clci,zhou2019d}. 

\section{Method}
\label{sec:method}
We propose a method to partition a 3D MRI volume $X \in \real^{C \times H \times W}$ into lesion (positive, 1) and normal (negative, 0) classes. Our method takes, as input, a 3D slice of $c$ consecutive 2D images $x \in \real^{c \times H \times W}$ ($c$ is an odd integer) from $X$ and predicts the binary segmentation for the image $\bar{x} \in \real^{1 \times H \times W}$, the $\frac{c+1}{2}$-th image of $x$. In other words, $x$ is a sliding window of $c$ images centered at a target image $\bar{x}$. To avoid sampling out of bounds, we perform mean padding of size $\frac{c-1}{2} \times H \times W$ on both sides of $X$ before sampling $x$ (see Sec. 1 of Supp. Mat. for more details). To segment a single image $\bar{x}$, we propose to learn a deep neural network $f_\omega$, parameterized by $\omega$, where $f : \real^{c \times H \times W} \mapsto [0, 1]^{1 \times H \times W}$ is a function that takes the 3D slice $x$ as an input and outputs the sigmoid response $f_\omega(x)$, a confidence map corresponding to lesions in $\bar{x}$. To obtain the binary segmentation of $X$, we aggregate our predictions by running $f_\omega$ for all $x$ and setting any response greater than a threshold of 0.5 to the lesion class. We note that our method can be extended to multi-class segmentation simply by expanding our output to $[0, 1]^{K \times H \times W}$ for $K$ classes, and choosing the class with highest response, i.e. $\arg \max f_\omega(\cdot)$, to yield the segmentation. 
\begin{figure}[t]
\centering
    \includegraphics[width=1.00\linewidth]{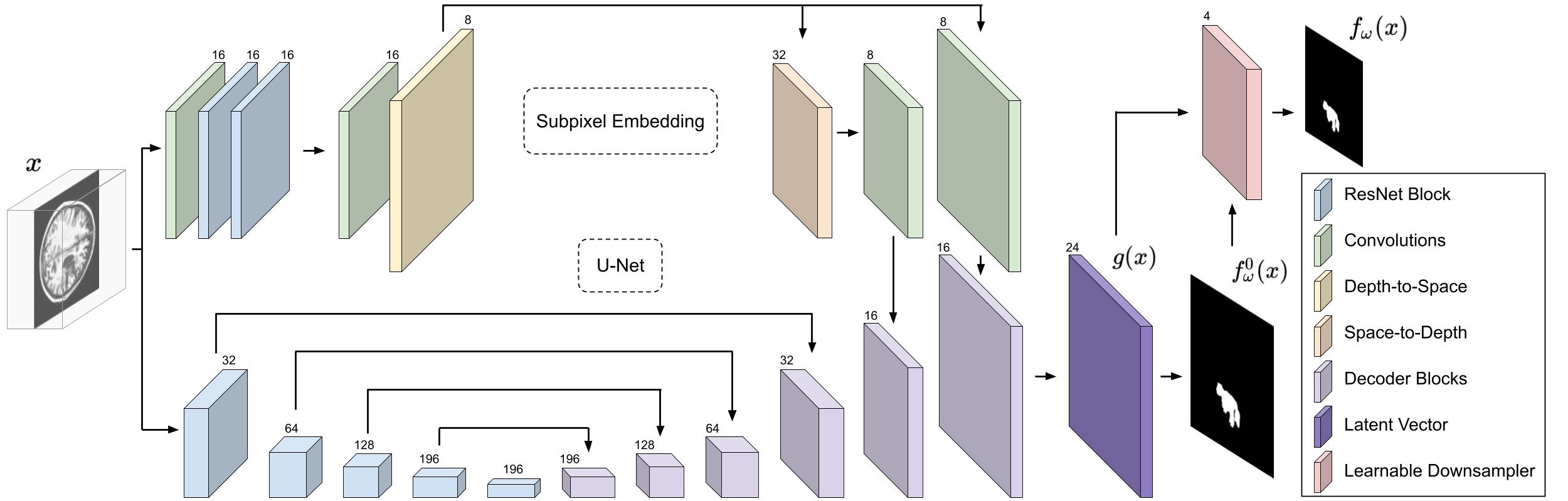}
    \caption{\textit{Network architecture.} \netname is comprised of (i) a U-Net based encoder-decoder that produces subpixel predictions $f^0_\omega (x)$ at $2\times$ the input resolution, which are guided by (ii) a subpixel embedding that captures local structure. The final output $f_\omega(x)$ is achieved by combining local predictions in a $2\times2$ neighborhood as a weighted sum based on the per element contribution predicted by a (iii) learnable downsampler.}
\vspace{-1em}
\label{fig:network_architecture}
\end{figure}

\subsection{Network Architecture}
\label{sec:network_architecture}
Our network $f_\omega$ (\figref{fig:network_architecture}) is composed of two modules: (i) an encoder-decoder (based on U-Net \cite{ronneberger2015u}) that outputs at $2 \times$ the input resolution, e.g. $2H \times 2W$, whose predictions are guided by (ii) a network that maps the input $x$ to a high dimensional embedding space also at twice the input resolution. The result is a confidence map comprised of ``subpixel'' predictions -- the output class for each input pixel is represented by four predictions within a $2 \times 2$ neighborhood. Rather than using hand-crafted downsampling techniques (e.g. bilinear, nearest neighbor) to obtain the output at the original ($1\times$) spatial resolution, we propose a learnable downsampler that predicts the weight, or contribution, of each subpixel prediction in a local region corresponding to the pixel in the $1\times$ resolution. For simplicity, we refer to our embedding function as a subpixel embedding and our overall architecture ($f_\omega$) as a subpixel network or ``SPiN'' for short (\figref{fig:network_architecture}). 

\begin{figure}[t]
\centering
    \includegraphics[width=1.00\linewidth]{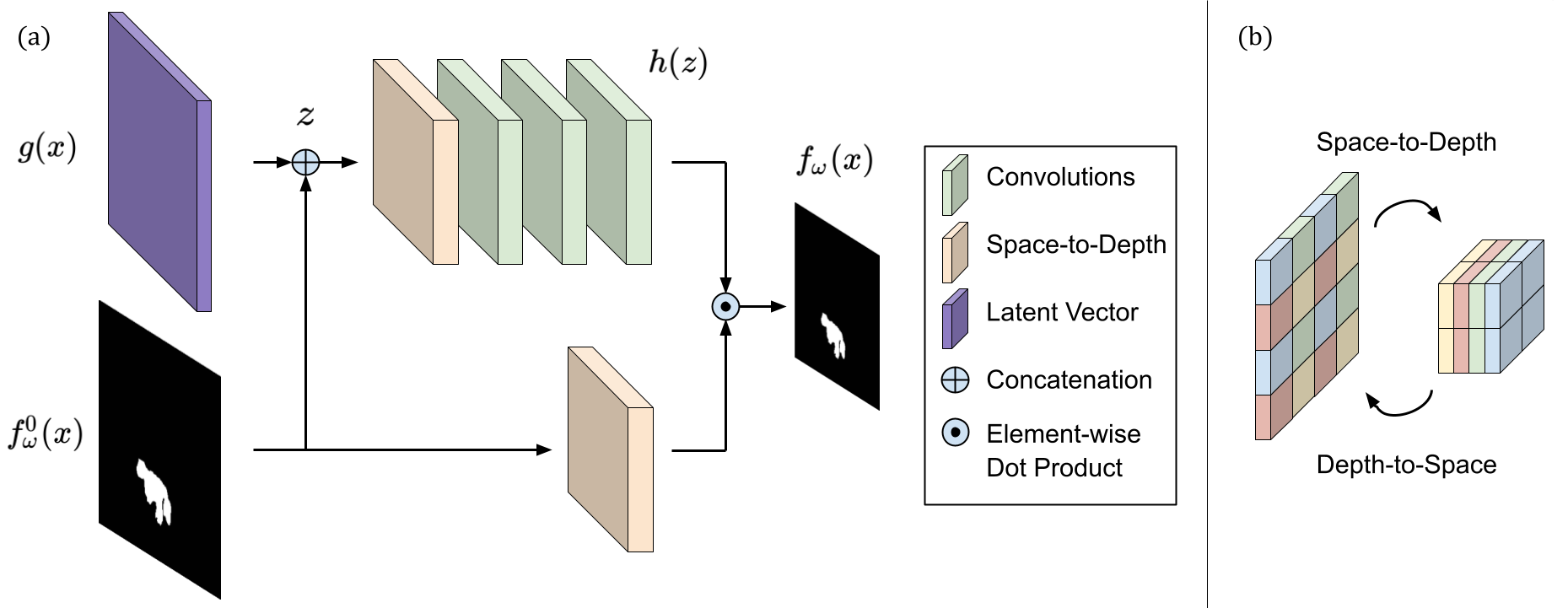}
    \caption{\textit{Learnable Downsampler, Space-to-Depth and Depth-to-Space}. (a): Learnable Downsampler predicts the contribution $h(z)$ of each subpixel prediction in $f_\omega^0(x)$ by conditioning on $f_\omega^0(x)$ and the latent vector $g(x)$. Subpixel predictions $f_\omega^0(x)$ are rearranged to the resolution of the input using Space-to-Depth. The final output $f_\omega(x)$ is produced by taking the element-wise dot product between $h(z)$ and the reshaped $f_\omega^0(x)$. (b) Space-to-Depth reduces resolution by rearranging elements from the spatial dimensions into the channel dimensions, where each $2 \times 2$ neighborhood is reshaped to a 4 element vector. Depth-to-Space conversely performs spatial expansion by rearranging elements from the channel dimensions to height and width dimensions.
    }
\vspace{-1em}
\label{fig:learnable_downsampler}
\end{figure}

\textbf{Subpixel Embedding} consists of feature extraction and spatial expansion phases. Feature extraction is performed by two ResNet blocks \cite{he2016deep} with 16 filters per layer; we also use stride of 1 and zero-padded edges to minimize spatial reduction. The extracted $16 \times H \times W$ feature maps are fed to a depth-to-space module  \cite{shi2016real} that rearranges elements from the channel dimension to the height and width dimensions (see \figref{fig:learnable_downsampler}-(b)). The resulting set of $4 \times 2H \times 2W$ feature maps with twice the spatial resolution then undergoes a $1 \times 1$ and a $3 \times 3$ convolution layers, with 8 filters each. The resulting $8 \times 2H \times 2W$ high dimensional feature maps, produced by our subpixel embedding function, resolve fine local details by increasing the feature map resolution and thus representing information at each pixel location with four ``subpixel'' feature vectors.

When used as skip connections, these embeddings complement the standard U-Net architecture that obtains a global representation of the input by spatial downsampling (striding and max pooling), which naturally discards local detail. Hence, we propose to inject these embeddings into the decoder via feature concatenation at the original ($1 \times$) resolution and at the $2 \times$ output resolution. To reduce the height and width dimensions of the embeddings to match the feature maps at the $1 \times$ resolution, we propose a space-to-depth module, which performs the inverse operation of depth-to-space (see \figref{fig:learnable_downsampler}-(b)), yielding $32 \times H \times W$ feature maps . Unlike striding and pooling, the depth-to-space operation is information preserving as it rearranges feature vectors from the height and width dimensions to their channel dimension. The result is fed through a $3 \times 3$ convolutional layer with 8 filters and concatenated with the feature maps of the decoder at the $1\times$ resolution. Similarly, the embeddings at $2\times$ resolution undergo a separate $3 \times 3$ convolution to yield the output resolution guidance before being concatenated with their corresponding feature maps in the decoder. 
Finally, the $2\times$ decoder output $f_\omega^0(x) \in [0, 1]^{1 \times 2H \times 2W}$ is produced by convolving a single $3 \times 3$ filter over the resulting latent vector $g(x) \in \real^{24 \times 2H \times 2W}$. We use subpixel guidance (SPG) to refer to the process of learning and injecting the embedding as skip connections, which substantially helps with localizing small lesions missed by previous works \cite{qi2019x,ronneberger2015u,yang2019clci,zhou2019d} (see \figref{fig:results_atlas_small_lesions}). We note that SPG is light-weight and only uses 16K parameters.

\textbf{Learnable downsampler} 
takes the concatenation $z = [g(x); f_\omega^0(x)]$ of the latent vector $g(x)$ and the $2\times$ resolution output $f_\omega^0(x)$ and predicts $h(z)$, where $h: \real^{25 \times 2H \times 2W}~\mapsto~[0, 1]^{4 \times H \times W}$. In other words, $h(z)$ is a set of $4 \times H \times W$ values that determine the contribution of each subpixel prediction in a $2 \times 2$ neighborhood of $f_\omega^0(x)$. To achieve this, we first perform space-to-depth on $z$ to rearrange each $2 \times 2$ neighborhood into a 4 element vector. This is followed by two $3 \times 3$ convolutions of 16 filters and a $1 \times 1$ convolution with 4 filters. $h(z)$ is the softmax response of the result along the channel dimension. 

To obtain the final output $f_\omega(x)$, we utilize space-to-depth to rearrange $f_\omega^0(x)$ into the shape of $4 \times H \times W$ (to match the shape of $h(z)$) and take its element-wise dot product with $h(z)$. With an abuse of notation, $f_\omega(x) = f_\omega^0(x) \cdot h(z)$. Because $h(z)$ is conditioned on the latent vector $g(x)$ of the input, the predicted weights respect lesion boundaries to yield detailed segmentations. This is unlike bilinear or nearest-neighbor downsampling where  weights are predetermined and independent of the input. We note that our learnable downsampler is also lightweight and only consists of 11K parameters. 

\subsection{Loss Function}
\label{sec:loss_function}
We assume a training set of $\{(x^{(n)}, \bar{y}^{(n)})\}_{n=1}^N$, where $\bar{y}^{(n)}$ is the ground truth corresponding to $\bar{x}^{(n)}$, the image located at the center of $x^{(n)}$. To train SPiN, we minimize the standard binary cross entropy loss,
\begin{equation}
    \ell(y, \bar{y}) = \frac{1}{|\Omega|}\sum_{u \in \Omega} -\big( \bar{y}(u) \log y(u) + (1 - \bar{y}(u))  \log (1 - y(u)) \big),
\end{equation}
where $\Omega \subset \real^2$ denotes the spatial image domain, $u$ a pixel coordinate, and $y = f_\omega(x)$ the network output. The loss over the training set of $N$ samples reads
\begin{equation}
    L(\omega) = \frac{1}{N} \sum_{n=1}^N \ell(f_\omega(x^{(n)}), \bar{y}^{(n)})).
\end{equation}
We note that previous works \cite{yang2019clci,zhou2019d} used soft Dice loss (an approximation of the true Dice score) to counter the class imbalance between normal and lesion tissues, characteristic in the lesion segmentation problem. However, a minimizer of cross entropy equivalently minimizes Dice, and empirically, we found that directly minimizing cross entropy yields better performance for our model. We hypothesize that our SPG allows small lesions to be recovered more easily, making our method more conducive to minimizing cross entropy, which is not prone to the noisy training signal inherent in soft Dice. We demonstrate this in row 7 of \tabref{tab:results_atlas_ablation} in our ablation studies. Also, we note that our loss can be easily extended for multi-class classification to accommodate multiple lesion categories.

\section{Experiments and Results}
\label{sec:experiments}
\begin{figure}[t]
\centering
    \includegraphics[width=1.00\linewidth]{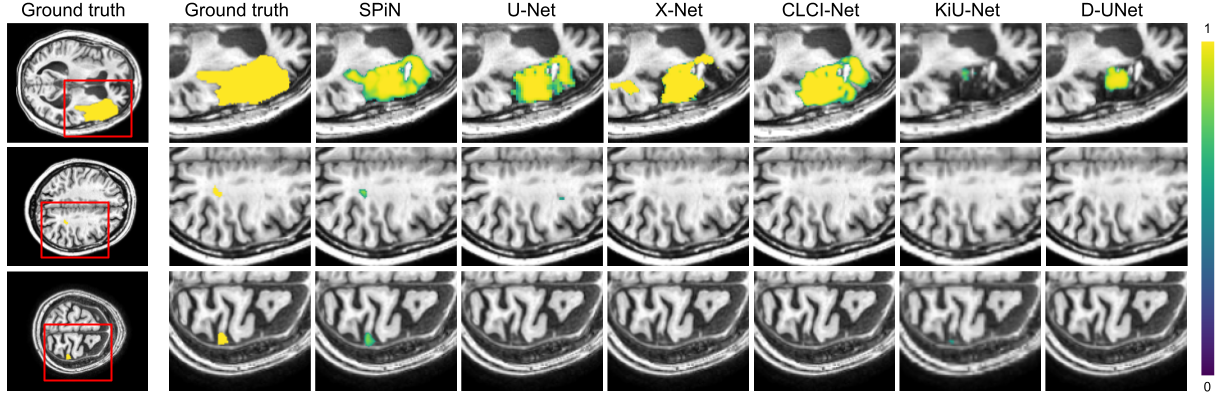}
    \vspace{-2.5em}
    \caption{\textit{Qualitative results on ATLAS.} Columns 2-8 show (zoomed in) head-to-head comparisons across all methods for highlighted areas in column 1. Row 1 demonstrates that \netname outperforms existing works in capturing shape and boundary details in medium-sized, irregularly-shaped lesions. Furthermore, rows 2 and 3 demonstrate SPiN's ability to localize small lesions that are missed by other models.}
    \vspace{-1em}
\label{fig:results_atlas_small_lesions}
\end{figure}

We demonstrate our method on the Anatomical Tracings of Lesion After Stroke (ATLAS) MRI dataset \cite{liew2018large,liew2017anatomical}, using the metrics defined in \tabref{tab:eval_metrics}. ATLAS contains 304 T1-weighted MRI scans of stroke patients with corresponding lesion annotations. The data is collected from 11 research sites worldwide, manually annotated, and post-processed (i.e. smoothing and defacing for privacy), leaving 239 patient scans with  189 2D images ($197 \times 233$ resolution) each. 
Since no official data split is provided by \cite{liew2018large}, previous works \cite{qi2019x,yang2019clci,zhou2019d} evaluated their methods using $k$-fold cross validation and randomly sampled data splits. However, the value of $k$ and samples within each split varied across works. Due to the lack of consistency, the reported results are not directly comparable. Thus, we propose a training (212 patients) and a held-out testing (27 patients) split to standardize the evaluation protocol for more rigorous comparisons. 
We provide quantitative comparisons against \cite{qi2019x,ronneberger2015u,valanarasu2020kiu,yang2019clci,zhou2019d} on the proposed training and testing split in \tabref{tab:results_atlas_trainval}. We also show qualitative (\figref{fig:results_atlas_small_lesions}) and quantitative (\tabref{tab:results_atlas_small_lesions}) comparisons on segmenting small lesions using a subset of test set: 490 images containing \textit{only} lesions smaller than 100 pixels (0.2\% of the image). All reported results for previous works are obtained using their training procedures and open-sourced code. We also provide details on our training and testing split in Sec. 2 of Supp. Mat. and further $k$-fold cross validation comparisons in Sec. 3. of Supp. Mat.

\textbf{Implementation details.} Our model is implemented in PyTorch~\cite{paszke2019pytorch} and optimized using Adam \cite{kingma2015adam}. We used an initial learning rate of $3 \times 10^{-4}$, decreased it to $1 \times 10^{-4}$ after 400 epochs, and to $5 \times 10^{-5}$ after 1400 epochs for a total of 1600 epochs. We choose $c = 5$ for the number images in the input $x$. During training, $\bar{x}$ and its corresponding $x$ are randomly sampled from $X$. Training takes $\approx8$ hours on an Nvidia GTX 1080 GPU, and inference takes $\approx11$ ms per 2D image. For data augmentation, we randomly perform (i) horizontal and vertical flips, (ii) rotation between -30$^{\circ}$ and 30$^{\circ}$, and (iii) add zero-mean Gaussian noise with standard deviation of $1 \times 10^{-2}$ to training samples. We perform augmentation with a probability of 1 for 1400 epochs and decrease it to 0.5 thereafter so training samples will be closer to the true distribution of the dataset.

\begin{table}[t]
\centering
    \setlength\tabcolsep{14.5pt}
    \begin{tabular}{l c c c c}
        \midrule
        Metric & IOU & DSC & Precision & Recall \\
        \midrule
        Definition 
        & $\frac{\textrm{TP}}{\textrm{TP} + \textrm{FN} + \textrm{FP}}$ 
        & $\frac{2 \times \textrm{TP}}{2 \times \textrm{TP} + \textrm{FN} + \textrm{FP}}$
        & $\frac{\textrm{TP}}{\textrm{TP} + \textrm{FP}}$
        & $\frac{\textrm{TP}}{\textrm{TP} + \textrm{FN}}$ \\
        \midrule
    \end{tabular}
    \caption{\textit{Evaluation metrics}. IOU denotes Intersection Over Union, and DSC denotes Dice similarity coefficient. $\textrm{TP}$, $\textrm{FN}$ and $\textrm{FP}$ correspond to true positive, false negative and false positive respectively.}
    \vspace{-1em}
\label{tab:eval_metrics}
\end{table}

\begin{table}[th]
\centering
\scriptsize
\setlength\tabcolsep{5.5pt}
    \begin{tabular}{l c c c c c c c c}
        \midrule
        & \multicolumn{4}{c}{Performance  Metrics} & & & \multicolumn{2}{c}{Memory Usage (GB)} \\
        \cmidrule{2-5}
        \cmidrule{8-9}
        Method & DSC & IOU & Precision & Recall 
        & Runtime (s) & & Train & Test \\
        \midrule
        U-Net \cite{ronneberger2015u}
        & 0.584 & 0.432 & 0.674 & 0.558 & 1.375 & & 2.291 & 1.181 \\ 
        D-UNet \cite{zhou2019d}
        & 0.548 & 0.404 & 0.652 & 0.521 & 3.425 & & 15.426 & 15.426 \\ 
        CLCI-Net \cite{yang2019clci}
        & 0.599 & 0.469 & 0.741 & 0.536 & 8.860 & & 7.853 & 7.853 \\ 
        KiU-Net \cite{valanarasu2020kiu}
        & 0.524 & 0.387 & 0.703 & 0.459 & 1.05 & & 23.566 & 1.555 \\
        X-Net \cite{qi2019x}
        & 0.639 & 0.495 & 0.746 & 0.588 & 5.046 & & 11.839 & 11.839 \\
        \netname (Ours)
        & \textbf{0.703} & \textbf{0.556} & \textbf{0.806} & \textbf{0.654}
        & 2.145 & & 3.273 & 0.803 \\
        \midrule
    \end{tabular}
    \caption{\textit{Quantitative comparison on ATLAS.} \netname outperforms all methods across all performance metrics. It is also one of the least computationally expensive models, i.e. smallest test time memory footprint, second in training memory usage, and third fastest in runtime per patient (189 images).}
\vspace{-1em}
\label{tab:results_atlas_trainval}
\end{table}

\textbf{ATLAS test set.} \tabref{tab:results_atlas_trainval} shows that our approach outperforms competing methods \cite{qi2019x,ronneberger2015u,valanarasu2020kiu,yang2019clci,zhou2019d} across all evaluation metrics. Specifically, we beat the best performing method X-Net \cite{qi2019x} by an average of $\approx10.4$\% with a 72.3\% reduction in training memory and a 57.5\% runtime reduction during inference. Our approach also uses a smaller memory footprint, containing only $\approx5.3$M parameters, compred to $\approx15$M in \cite{qi2019x}. Another key comparison is with KiU-Net, which learns a representation at $8\times$ the original input spatial resolution.
Unlike us, KiU-Net \cite{valanarasu2020kiu} uses max pooling layers, which discards information, to reduce the size of their high resolution representation to the original ($1\times$) resolution. Whereas, we maintain the $2\times$ resolution of our embedding until the output layer, which yields subpixel predictions that are aggregated by our learnable downsampler to the $1\times$ resolution. 
Admittedly, this comes at the cost of runtime -- our method requires 2.145s per patient and KiU-Net \cite{valanarasu2020kiu} requires 1.05s. However, we outperform \cite{valanarasu2020kiu} by an average of $33.7$\% across all metrics and reduce test time memory by half. 
We show qualitative comparisons in row 1 of \figref{fig:results_atlas_small_lesions} where the segmentation produced by our approach better captures irregularly shaped lesions than those predicted by competing methods. 

\begin{table}[ht]
    \centering
    \scriptsize
    \setlength\tabcolsep{20.2pt}
    \begin{tabular}{l c c c c}
        \midrule
        Method & DSC & IOU & Precision & Recall \\
        \midrule
        U-Net \cite{ronneberger2015u}
        & 0.368	& 0.225	& 0.440	& 0.316 \\ 
        D-UNet \cite{zhou2019d}
        & 0.265 & 0.180 & 0.377 & 0.264 \\
        CLCI-Net \cite{yang2019clci}
        & 0.246 & 0.178 & \textbf{0.662} & 0.215 \\
        KiU-Net \cite{valanarasu2020kiu}
        & 0.246 & 0.255 & 0.466 & 0.206 \\
        X-Net \cite{qi2019x}
        & 0.306 & 0.213 & 0.546 & 0.268 \\
        \netname (Ours)
        & \textbf{0.424} & \textbf{0.269} & 0.546 & \textbf{0.347} \\
        \midrule
    \end{tabular}
    \caption{\textit{Evaluation on small lesion subset}. While \cite{yang2019clci} achieves the highest precision, we note they have the second lowest recall out of all methods -- missing small lesions can negatively impact patient recovery. In contrast, our method ranks second in precision and first across all other metrics.}
    \vspace{-1.8em}
\label{tab:results_atlas_small_lesions}
\end{table}

\begin{table}[t]
\centering
    \scriptsize
    \setlength\tabcolsep{12pt}
    \begin{tabular}{l c c c c}
        \midrule
        Method & DSC & IOU & Precision & Recall \\
        \midrule
        Without SPG, LD (Baseline)
        & 0.634	& 0.487	& 0.707	& 0.606 \\ 
        Without SPG
        & 0.637 & 0.487 & 0.701 & 0.613 \\
        Replace SPG with addit. convolutions 
        & 0.627 & 0.475 & 0.721 & 0.596 \\
        Replace SPG w/ bilinear upsampling
        & 0.663 & 0.513 & 0.780 & 0.600 \\
        Replace SPG w/ nearest upsampling
        & 0.660 & 0.513 & 0.762 & 0.626 \\
        Replace LD with downsampling  
        & 0.670 & 0.526 & 0.786 & 0.625 \\
        Full model with soft Dice loss
        & 0.684 & 0.546 & 0.729 & 0.672 \\
        Full model
        & \textbf{0.703} & \textbf{0.556} & \textbf{0.806} & \textbf{0.654} \\
        \midrule
    \end{tabular}
    \caption{\textit{Ablation study on ATLAS}. Removing SPG and/or LD results in performance decrease (rows 1, 2, 6), and SPG cannot be substituted with more parameters or interpolation (rows 3-5). The best results are achieved by our full model (row 8).}
    \vspace{-1em}
\label{tab:results_atlas_ablation}
\end{table}

\textbf{Small lesion segmentation.} Here, we consider the task of segmenting lesions that occupy fewer than 100 pixels or 0.2\% of the image. Due to the challenging nature of the task, we observe an expected drop in performance across all methods (trained on the proposed split) when segmenting small lesions (\tabref{tab:results_atlas_small_lesions}), as compared to doing so for all lesion sizes (\tabref{tab:results_atlas_trainval}). However, we still outperform all competing methods -- by even larger margins than on the full test set. This shows that competing methods, while able to localize large and medium sized lesions, actually perform poorly on small lesions. With the exception of precision, where we tie for second with X-Net \cite{qi2019x}, we rank first in all other metrics. We note that while CLCI-Net \cite{yang2019clci} has the highest precision, it also achieved second lowest recall, meaning that it misses many small lesions, which is critical to clinical prognosis and thus patient recovery. This is also reflected in DSC and IOU where we outperform \cite{yang2019clci} by 72\% and 51\%, respectively. Qualitatively, rows 2 and 3 in \figref{fig:results_atlas_small_lesions} show that our method successfully localized small lesions that \cite{qi2019x,ronneberger2015u,valanarasu2020kiu,yang2019clci,zhou2019d} missed entirely.

\textbf{Ablation studies.} \tabref{tab:results_atlas_ablation} shows the effect of each of our contributions to architectural design. Row 1 shows that our baseline, a U-Net \cite{ronneberger2015u} based encoder-decoder, performs significantly worse by $11.7\%$ than the proposed approach because it lacks fine local details from SPG and uses bilinear downsampling instead of a learnable downsampler (LD). Including LD alone, but not SPG (row 2) provides no improvement as the network only learns a coarse global representation, but is still missing details lost during spatial downsampling.

In row 3, we show that solely increasing parameters (i.e. adding ResNet blocks \cite{he2016deep} to the baseline) brings no improvement, which suggests that the performance boost is \textit{not} a result of a larger network. In fact, SPG and the learnable downsampler marginally increase the model size as they only combine for 27K parameters. Rows 4 and 5 show that using hand-crafted $2\times$ resolution images (from bilinear, nearest neighbor upsampling) does provide some gain. In these experiments, we replace SPG with different interpolation methods and the higher resolution images undergo $3 \times 3$ convolutions before being passed as skip connections to the decoder. However, because the $2\times$ representation is not learned, as it is with SPG, the result is still $\approx6\%$ worse than our full model. 
Our learnable downsampler (LD) contributes $4.4\%$ to our performance (row 6) as removing LD and replacing it with bilinear interpolation smooths lesion boundaries, resulting in loss of details. Finally, we justify the use of cross entropy for our loss function; row 7 demonstrates that minimizing a soft Dice loss, as in \cite{yang2019clci,zhou2019d}, results in worse performance. The best performance is achieved with our full model using SPG and LD, and minimizing cross entropy (row 8).

\section{Discussion}
\label{sec:discussion}
We propose SPiN, a network architecture that learns a spatially increasing embedding that, when used as guidance for an encoder-decoder network, helps ensure that small structures are not lost through spatial downsampling in the encoder. We note that our embedding does not create extra spatial information (data processing inequality), but serves as a means for better characterization of local regions for the downstream segmentation task. While we outperform existing works and improve on small lesion segmentation, we do cost more memory and compute than the baseline. However, the extra cost is within reason (1 GB of memory for training and $\approx$ 0.7s in runtime) and does not limit applicability. Despite the improved segmentation performance, we would like to address that there is still room for improvement, especially with small lesions. The highest recall of 0.347 (\tabref{tab:results_atlas_small_lesions}) achieved by our model is admittedly low compared to recall metrics on the full dataset, implying that many small lesions still pass undetected. We note that this is one of the first works to study subpixel architectures in lesion segmentation, and we hope that our optimistic results will motivate further exploration in this direction.

\textbf{Acknowledgements.} This work was supported by NIH-NEI 5R01EY029689 and R01EY030595, ARO W911NF-17-1-0304, NRF-2017R1A2B4006023 and IITP-2021-0-01341, AI Graduate School (CAU) in Korea.

\bibliographystyle{splncs04}
\bibliography{ref}

\newpage
\vspace*{0.5cm}
\begin{center}
    \Large{
        \textbf{
            Small Lesion Segmentation in Brain MRIs with Subpixel Embedding \\
            \vspace{1cm}
            SUPPLEMENTARY MATERIALS
        }
    }
    \vspace{1cm}
\end{center}

\appendix

\noindent \textbf{Summary of content:} In \secref{sec:selecting_slices}, we elaborate on how we select the input $x$ from the 3D MRI scan during training and testing phases. In \secref{sec:details_atlas_train_test_split}, we discuss details on our proposed training and testing splits and show their data distribution with respect to lesion size. \secref{sec:results_atlas_crossval} provides comparisons between all competing methods using the reported numbers from their papers. We note that these numbers are not directly comparable since each method used a different data split. In \secref{sec:small_lesion_visualization}, we provide additional qualitative examples of our predictions on MRI images for the task of small lesion segmentation. Finally, we show the feature maps produced by our subpixel embedding in \secref{sec:feature_maps_subpixel_embedding} and discuss how they contribute to guiding our decoder in localizing small lesions and recovering irregularly shaped lesion boundaries.

\begin{figure}[th]
\centering
    \includegraphics[width=1\linewidth]{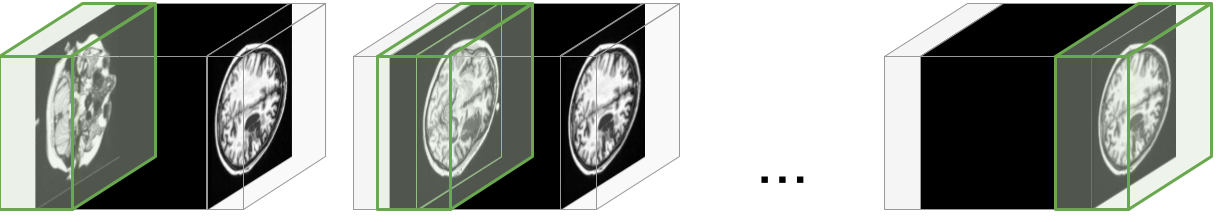}
    \caption{\textit{Selecting slices.} First, a 3D MRI scan $X \in \real^{C \times H \times W}$ is mean padded on either side along its channel dimension. The mean padding is represented by the appended gray boxes on either side of the grayscale MRI scan. The green boxes represent a selected volume $x \in \real^{c \times H \times W}$ consisting of $c$ 2D images centered at $\bar{x}$. In the leftmost figure, $\bar{x}$ is selected to be in one end of the 3D scan $X$, so $x$ will contain mean padding images. We select $\bar{x}$ in a sliding window fashion until the right most figure where $\bar{x}$ is chosen to be in the other side of $X$.}
\vspace{-1em}
\label{fig:selecting_slices}
\end{figure}

\section{Selecting Slices from a 3D MRI}
\label{sec:selecting_slices}
In the Sec. 3 of the main text, we discussed how we select $c$ consecutive 2D images from the 3D MRI; here we will elaborate on this. Recall that $c$ is an odd integer. First, each MRI $X \in \real^{C \times H \times W}$ is padded with the mean of the entire dataset $\frac{c-1}{2}$ times along the first dimension, $C$, parallel to the axial plane. The resulting tensor $X'$ has a shape of $(C+c-1) \times H \times W$.  Next, we select a subset of this volume, as seen in green in \figref{fig:selecting_slices}, consisting of $c$ 2D images centered at the image $\bar{x} \in X'$ such that the condition $\bar{x} \in X$ is also satisfied. Thus, $\bar{x}$ is a 2D image of the \textit{original} MRI scan, and it is for $\bar{x}$ that we make our segmentation prediction. Besides $\bar{x}$, the volume is filled with neighboring images or the mean-padded values when no neighbors are available (i.e. near the edges of the 3D MRI scan), allowing us to utilize contextual information without reducing the amount of data. With an abuse of notation and following the notation used in the main text, we will refer to $X'$ as $X$.

During training, $\bar{x}$ and its corresponding $x$ is randomly sampled from $X$. For the stability of training, we also increase the probability of sampling $\bar{x}$ with lesion of any size to 95\%. This is done to counteract the class imbalance where images without lesions greatly outnumber those with lesions. During test time, we iteratively select $\bar{x}$ and $x$ from $X$, starting from image 1 to $C$ for a total of $C$ images in $X$. For the size of $x$, we chose $c$ based on hyper-parameter tuning. We note that performance does not significantly improve if we choose $c$ to be larger than 5; however, choosing a large $c$ (such as up to the entire 3D volume like methods with 3D architectural backbones \cite{hui2020partitioning,zhou2019d}) will increase memory usage and computation requirements, rendering the method impractical. We also note that choosing $c=1$ will reduce performance by $\approx 5\%$ on average.

\begin{figure}[t]
\centering
    \includegraphics[width=0.49\linewidth]{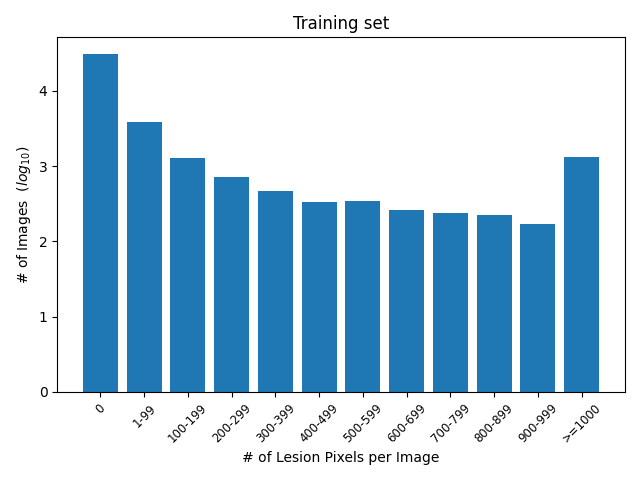}
    \includegraphics[width=0.49\linewidth]{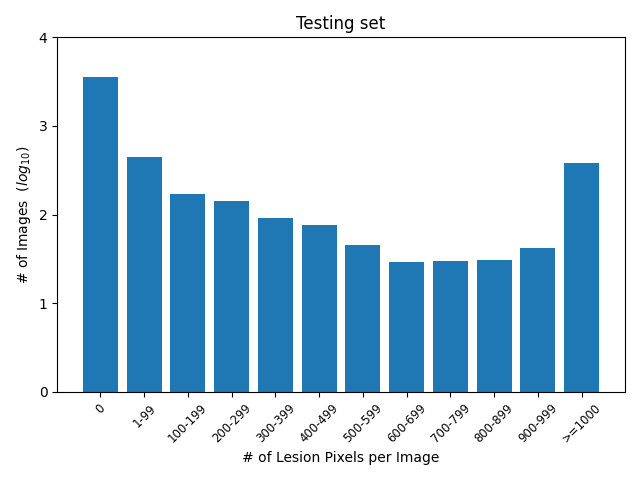}
    \caption{\textit{Distribution of lesions in the proposed training and testing set for ATLAS.} Number of images (samples) are shown in $\log$ scale. We were careful to select the testing set (right) such that it is representative of the training set (left) distribution. Also we ensured that there are sufficient small lesions (fewer than 100 pixels) in the testing set to make the dataset challenging. The small lesion subset is represented by the second bar (1 - 99) in the testing set plot on the right.}
\vspace{-1em}
\label{fig:atlas_train_test_split_distro}
\end{figure}

\section{Details on ATLAS Training and Testing Split}
\label{sec:details_atlas_train_test_split}
In Sec. 4 of the main text, we proposed a training and testing split for the ATLAS \cite{liew2018large,liew2017anatomical} dataset consisting of 212 patients for training (40K images) and a held-out testing set of 27 patients (5.1K images). In \figref{fig:atlas_train_test_split_distro} we show the distribution of lesion sizes for the training set (left) and the testing set (right). The data split was randomly chosen and post-processed to ensure that the testing set is representative of the training set data distribution. To make sure that the benchmark was challenging, we  ensured that there are sufficient images with small lesions in the testing set. The proposed small lesion evaluation uses the subset of the testing set that contain \textit{only} lesions with fewer than 100 pixels. Both the training and testing split and the small lesion subset will be released along with our open-sourced code and pretrained models upon publication. We cordially invite researchers to test on this benchmark in order to push the state of the art for small lesion segmentation in MRIs.

\begin{table}[t]
\centering
    \setlength\tabcolsep{18.5pt}
    \begin{tabular}{l c c c c}
        \midrule
        Method & DSC & IOU & Precision & Recall \\
        \midrule
        U-Net \cite{ronneberger2015u} 
        & 0.475 & 0.358 & 0.586 & 0.471 \\
        D-UNet \cite{zhou2019d} 
        & 0.535 & - & 0.633 & 0.524  \\
        CLCINet \cite{yang2019clci} 
        & 0.581	& - & 0.649 & 0.581 \\
        KiU-Net \cite{valanarasu2020kiu}
        & 0.441 & 0.321 & 0.580 & 0.418 \\
        X-Net \cite{qi2019x} 
        & 0.487 & 0.372 & 0.600 & 0.475 \\ 
        SPiN (Ours)
        & \textbf{0.587} & \textbf{0.461} & \textbf{0.670} & \textbf{0.610} \\
        \midrule
    \end{tabular}
    \caption{\textit{Self reported results on ATLAS}. \cite{yang2019clci,zhou2019d} did not report on IOU. Results for \cite{ronneberger2015u} cross validation were reported by \cite{qi2019x}. Because \cite{valanarasu2020kiu} did not evaluate on ATLAS, we used the open-sourced code provided by the authors to train their model on our cross validation split. We report results obtained by training their model based on the procedure specified in their paper.}
\vspace{-1em}
\label{tab:results_atlas_crossval}
\end{table}

\section{Comparison of Reported Results on ATLAS}
\label{sec:results_atlas_crossval}
As noted above and in Sec. 4 of the main text, there is not a clear precedent of evaluating and comparing related works. Due to the lack of consistency, we proposed a training and testing split on which we trained and evaluated \cite{qi2019x,ronneberger2015u,valanarasu2020kiu,yang2019clci,zhou2019d} in our main text. For fairness, we directly used the open-sourced code provided by the authors and trained their model on our proposed data split. The results in Table 2 and 3 in the main text were obtained by training their model using the procedure specified in their papers. In addition, to account for random seed affecting performance, we trained each of their models multiple times and selected the best performing checkpoint for evaluation.

For completeness, we also compare our method to reported results of other works in \tabref{tab:results_atlas_crossval}. Our reported results were conducted on 6-fold cross validation to minimize the impact of overfitting or selection bias on our results and better measure generalizability. Although these results seem to demonstrate that our model outperforms all other works, we note that the reported numbers from each method were obtained from training on different data splits. Hence, we  do not feel these numbers are directly comparable and refer the reader to the Table 2 and 3 in the main text for more rigorous comparisons.

\begin{figure}[t]
\centering
    \includegraphics[width=1\linewidth]{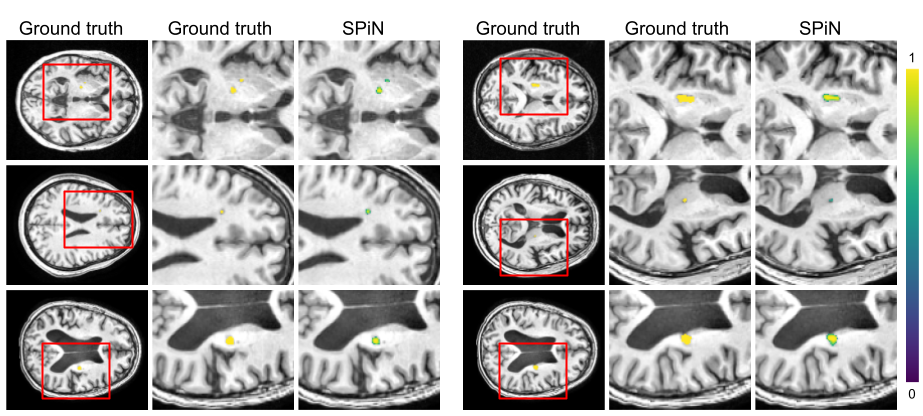}
    \caption{\textit{Qualitative results on small lesions on ATLAS.} Here we further demonstrate SPiN's performance specifically for small lesions (no larger than 100 pixels). In each $1 \times 3$ panel, the leftmost column shows the entire patient scan with the ground truth overlayed, the middle column shows the zoomed in binary ground truth in yellow, and the rightmost column shows SPiN's zoomed in softmax prediction. Our method is able to segment very small lesions that may otherwise be missed by other methods.}
\vspace{-0.5em}
\label{fig:qual_res}
\end{figure}

\section{Small Lesion Prediction Visualization for SPiN}
\label{sec:small_lesion_visualization}
We presents qualitative results of our model on additional small lesion test samples in \figref{fig:qual_res}. Identifying small lesions is a challenging task as the lesion area is less than 0.2\% of the image and requires the model to capture fine-grained details. As noted in the main text, accurately segmenting lesions as early as possible has a significant impact on patient recovery, yet most current works are unable to detect the smaller lesions\footnote{Early onset of stroke tend to yield smaller lesion sizes, which grow to maximum volume after 5 to 7 days \cite{beaulieu1999longitudinal,brott1989measurements,pantano1999delayed}; hence, the inability to detect small lesion can be detrimental to patient recovery.}. Here, we demonstrate that our model can segment lesions as small as a few pixels large, which in the time-sensitive context of stroke lesions could significantly affect treatment and recovery. 

\begin{figure}[t]
\centering
    \includegraphics[width=1\linewidth]{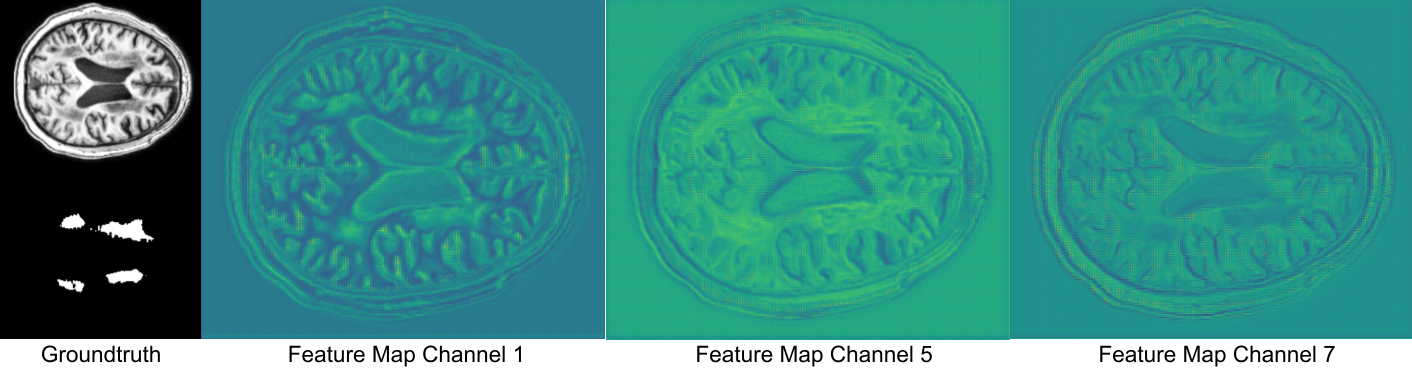}
    \caption{\textit{Feature maps from our subpixel embedding network.} Column 1: sample MRI image and corresponding ground truth. Columns 2-4: feature maps produced by our embedding network. Despite feeding $x \in \real^{c \times H \times W}$ into our embedding network, it naturally learns to  retain the structural details for $\bar{x}$, the target image to be segmented.}
\vspace{-1em}
\label{fig:feature_maps}
\end{figure}

\section{Features Maps from Subpixel Embedding}
\label{sec:feature_maps_subpixel_embedding}
In Sec. 3 of the main text, we discussed the architecture of our subpixel embedding network used for guiding the output of the decoder -- subpixel guidance (SPG). Here, we show examples of the feature maps produced by the learned embedding in \figref{fig:feature_maps}. Our subpixel embedding retains the anatomical structures in the output high resolution feature maps  that resembles those present in the low resolution MRI image. We note that this is particularly interesting because our only form of supervision is the cross entropy loss (Eqn. 1, 2 from main text) e.g. we do not use high resolution MRI images to constrain the latent representation of the subpixel embedding network, so the embedding does not necessarily need to learn shapes that resembles in image of interest. However, as seen in \figref{fig:feature_maps}, despite the input being $x \in \real^{c \times H \times W}$, which contains \textit{multiple images}, our embedding naturally learns to extract features that retain the structures of $\bar{x}$, the target image to be segmented. We hypothesize that the network is able to learn this without additional supervision because the target image is always located in the center of $x$.

As seen in \figref{fig:feature_maps}, the feature maps can be treated as a high dimensional super-resolved version of the input target image where small details are now represented by four times ($2 \times 2$ neighborhood) the number of elements. When used as guidance for the decoder via skip connections, the local structures preserved by these feature maps help the network resolve fine-grained details in the image for the localization of small lesion and the recovery of irregular lesion boundaries.

\end{document}